\documentclass{article}

\usepackage{arxiv}

\usepackage[utf8]{inputenc} 
\usepackage[T1]{fontenc}    
\usepackage{hyperref}       
\usepackage{url}            
\usepackage{booktabs}       
\usepackage{amsfonts}       
\usepackage{nicefrac}       
\usepackage{microtype}      
\usepackage{lipsum}		
\usepackage{graphicx}

\usepackage{doi}
\usepackage[backend=biber,style=apa]{biblatex}
\DeclareLanguageMapping{english}{english-apa}
\usepackage{appendix}       
\usepackage{pdfpages}       

\title{Comprehensive AI Assessment Framework: Enhancing Educational Evaluation with Ethical AI Integration}

\author{ \href{https://orcid.org/0000-0001-8846-7243}{\includegraphics[scale=0.06]{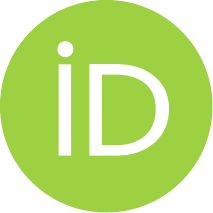}\hspace{1mm}Selçuk Kılınç}\thanks{Ph.D. in Science Education}\thanks {MS in Software Engineering \& Secondary Sci. \& Math. Edu.}\thanks{\href{https://scholar.google.com/citations?user=oghAi08AAAAJ&hl=tr}{Google Scholar} - \href{https://www.researchgate.net/profile/Selcuk-Kilinc-3}{ResearchGate} - \href{https://www.linkedin.com/in/selcukilinc/}{LinkedIn}} \\
	Mathematics and Science Education\\
	Middle East Technical University\\
	Ankara, Turkey \\
	\texttt{skilinc@metu.edu.tr} \\
}

\renewcommand{\headeright}{Kılınç, s.}
\renewcommand{\shorttitle}{\textit{Comprehensive AI Assessment Framework} }

\hypersetup{
pdftitle={Comprehensive AI Assessment Framework},
pdfsubject={ai.CS},
pdfauthor={Kılınç, S.},
pdfkeywords={Comprehensive AI Assessment Framework,AI Assessment, AI in Education, AI Integration, AI Ethics},
colorlinks=true,       
linkcolor=blue,        
citecolor=blue,        
filecolor=blue,     
urlcolor=blue          
}

\begin{document}
\maketitle

\begin{abstract}
	The integration of generative artificial intelligence (GenAI) tools into education has been a game-changer for teaching and assessment practices, bringing new opportunities, but also novel challenges which need to be dealt with. This paper presents the Comprehensive AI Assessment Framework (CAIAF), an evolved version of the AI Assessment Scale (AIAS) by Perkins, Furze, Roe, and MacVaugh, targeted toward the ethical integration of AI into educational assessments. This is where the CAIAF differs, as it incorporates stringent ethical guidelines, with clear distinctions based on educational levels, and advanced AI capabilities of real-time interactions and personalized assistance. The framework developed herein has a very intuitive use, mainly through the use of a color gradient that enhances the user-friendliness of the framework. Methodologically, the framework has been developed through the huge support of a thorough literature review and practical insight into the topic, becoming a dynamic tool to be used in different educational settings. The framework will ensure better learning outcomes, uphold academic integrity, and promote responsible use of AI, hence the need for this framework in modern educational practice.
\end{abstract}

\keywords{Comprehensive AI Assessment Framework   \and AI Assessment  \and AI in Education  \and AI Integration \and AI Ethics}

\section{Introduction}
Generative artificial intelligence tools, also known as GenAI tools, have had a transformative impact in numerous domains, including education \parencite[]{mello_education_2023}. Generation of human-like creative and problem-solving content for users has been made possible by the use of very advanced AI \parencite[]{yeo_artificial_2023}. GenAI tools are a tremendous development in AI technology. These tools autonomously create content that mimics human creative and problem-solving capabilities \parencite[]{dickey_model_2024}. Examples include ChatGPT, a language model designed for generating human-like text; conversational agents Gemini and Copilot; vision-language models Midjourney and Dall-E, a transformer whose decoder is conditioned on text functioning as a text-to-image generator \parencite[]{wang_matgpt_2024}. The integration of GenAI tools in education has opened up new possibilities for both students and educators \parencite[]{bubeck_sparks_2023}.

GenAI tools have succeeded with all their instances as they all generate something in return: either a human-like conversation or a vision from text \parencite[]{zhang_one_2023}. Initially, the technology appeared with rule-based approaches and modest datasets in the very early AI systems, limiting their capabilities \parencite[]{gampala_is_2020}. Technological advancements in deep learning and neural networks have allowed the development of powerful GenAI tools. For instance, ChatGPT generates human-like text through meaningful conversations and creates coherent and contextually relevant conversations \parencite[]{boscardin_chatgpt_2024}.

GenAI tools have application areas in a wide range of disciplines. In medicine, AI has become prominent with diagnostic tools and wearable technology, both clinical- and patient-facing \parencite[]{yeo_artificial_2023}. The broader impact of AI in medical education can be observed in conversational models such as ChatGPT \parencite[]{boscardin_chatgpt_2024}. Moreover, GenAI tools has been implemented dramatically in the disciplines of language education and library services, and commercial markets and management \parencites{gao_language_2024}{pack_using_2024}. Both domains explore the recent research agenda of AI technology applications for digital transformation. Such applications have been developed to assist in the decision-making process of managerial functions, in the facilitation of operations, or in market competition \parencite[]{kitsios_artificial_2021}. 

In education, AI technologies are used in a variety of beneficial ways. Evidence demonstrates that social science and humanities programs may find AI tools, such as ChatGPT, to be valuable assets in teaching students the skills they need to engage with modern practices \parencite[]{simms_work_2024}. But along with the commencement of the use of such tools, a significant domain required to be addressed has emerged.

Ethical considerations will keep playing a crucial role in GenAI tools. Addressing the ethical and pedagogical dimensions and encouraging responsible AI practitioners to uphold ethical standards and best practices are critical \parencite[]{pack_using_2024}. Such technological tools have fundamentally modified aspects of AI. Through further exploration of AI tools, ethically responsible coordinators will comprehend the GenAI tools and their subsequent impacts on society, the educational aspects of the users, and the surrounding world \parencite[]{sullivanetal_2023}. Only in this way can AI be harnessed to make a positive difference and develop technologies for people.

\subsection{Impact on Education and Assessment}

GenAI tools have had a significant effect on education, changing how people teach and learn. According to \textcite{mahligawati_artificial_2023}, the course of teaching as a profession may be disrupted by GenAI tools. ChatGPT, Gemini, Dall-E or etc., means of creativity and engagement, may be used to increase student participation beyond lectures in higher education institutions. Opting to teach AI to students of higher education may be preferred compared to other subjects \parencite[]{adiguzel_revolutionizing_2023}. In education, AI can individualize the training, give immediate feedback on assignments, and create an interactive atmosphere for the learners \parencite[]{kilinc_embracing_2023}.

In terms of assessment methods, GenAI tools have changed the way of assessments’ performance through automated scoring systems, personalized feedback, and adaptive testing tailored to the needs of each student \parencite[]{olga_generative_2023}. \textcite{kamalov_new_2023} stated that assessment processes may be sped up using AI since it reduces biases while also providing assistance for the deep comprehension of performance levels amongst pupils. Using these tools, educationalists should design assessments reflecting 21st-century goals as well as accommodating various styles of learning \parencite[]{singh_impact_2022}. Educationalists have to match the assessment tasks with the curriculum outcomes by providing personalized feedback \parencite[]{sajja_artificial_2023}.

Attitudes towards GenAI tools aimed at education differ regionally among people. Among the educationalists of all educational levels, some find the GenAI tools useful in improving teaching methods and learning outcomes, while others have concerns about cheating during exams as well as ethics relevant to the use of such a technology \parencite[]{sullivanetal_2023}. Establishing supportive attitudes about the positive potentials associated with the integration of AI at schools should therefore take into account these diverse perspectives in order to address any potential barriers \parencite[]{park_implementing_2024}. 

When it comes to the education sector, the use of AI has led to many arguments. For instance, questions have been asked concerning AI’s pedagogical implications, ability to boost student engagement, and impact on the pursuit of academic honesty \parencite[]{zhang_ai_2021}. Moreover, with the use of this technology, educationalists are now able to determine how they can personalize each student’s learning and develop interactive experiences with the learners, in addition to its numerous other advantages \parencite[]{pack_using_2024}. 

Moreover, \textcite{simms_work_2024} pointed out the consideration of ethical issues in addition to the negatively affected traditional methods as a consequence of AI’s adoption and thus suggested full educationalist training before any further emergence of negative outcomes, stating that much more could have been performed in a different way if only the educationalists were more interested and concentrated at the meetings where people had drawn attention to great points. \textcite{lane_tool_2024} advised that the stakeholders have to keep a close eye on both sides while integrating GenAI tools as they are the best means of teaching the students.

In special needs education, AI promotes inclusive pedagogy and supports students with diverse learning needs \parencite[]{gard_sharma}. AI can help create personalized learning experiences for individuals, adjust teaching approaches, and cultivate an inclusive educational atmosphere that will improve learning outcomes as well as ensure fairness for all the students \parencite[]{maghsudi_personalized_2021}. This will be achieve through the use of technology.

In brief, the incorporation of GenAI tools in the domain of education has the potential to extensively change the way we teach and learn, the methods of assessment utilized, and student involvement. Obviously, there are many advantages brought about by such technologies; however, it is important for practitioners not to ignore issues concerning academic honesty, ethical application, or even educationalist training procedures during the utilization of the same. Therefore, stakeholders should investigate the risks posed by AI in education and testing so as to reveal its full potential and create innovative environments that are accessible to everyone.

\section{Reactions and Ineffectiveness Against GenAI Tools}

The use of AI in education has faced both deep approval and strong rejection. This part focuses on the initial adoption of GenAI tools. It also addresses the prohibitions and limitations of the adoption of AI detection devices in order to control their use.

\subsection{Initial Reactions: Bans and Restrictions}

The introduction of AI in schools has faced different reactions, with some schools opting to ban or limit its use. Such initial reactions were driven by privacy concerns, data security issues, and a fear that it could disrupt traditional teaching methods \parencite[]{volante_leveraging_2023}. However, research shows that these measures are hardly effective. According to \textcite{hong_data_2022}, such bans can easily be circumvented, thus making them unreliable.

While intending to protect student privacy as well as uphold academic integrity, such restrictions often lose the plot. By discouraging the use of AI tools through clear bans, schools risk missing out on the potential benefits brought about by their use, such as improving learning outcomes through formative assessment practices. Instead, educational institutions may integrate the AI tools in an ethically responsible manner that supports students’ learning and development \parencite[]{volante_leveraging_2023}.

The discussion around the regulation of AI goes beyond education and brings about more extensive ethical questions. \textcite{morley_ethics_2021} argued for the need for more pragmatic ethics in AI and emphasized the continual assessment of ethics involved in every stage, from designing the algorithms to deploying systems at businesses or government agencies, etc. Therefore, rather than adopting all inclusive bans, we should focus only on those areas of high risk while putting in place the necessary safeguards \parencite[]{de_laat_companies_2021}.

Moreover, the evolving nature of AI technology presents additional challenges for regulation. \textcite{lam_randomized_2022} noted the lack of user-friendly tools for creating interactive educational resources, highlighting a potential gap for effective AI education. Addressing these gaps is essential for fostering AI literacy among students and educationalists.

Eventually, while initial reactions to GenAI tools in education are marked by attempts to ban or restrict their use, such approaches have proven to be highly ineffective. A deliberate understanding of the benefits and challenges associated with the integration of AI is necessary to develop more effective regulatory frameworks and educational practices.

\subsection{AI Detection Tools: Working Principle and Limitations}

The use of AI detection tools has become a common practice for managing the integration of GenAI tools in education. These tools employ machine learning algorithms, neural networks, and deep learning techniques to identify patterns and anomalies in data \parencite[]{adiguzel_revolutionizing_2023}. Despite their advanced capabilities, AI detection tools have notable limitations.
AI detection tools analyze text for patterns typically associated with AI-generated content, such as uniform sentence structure, specific word overuse, and predictable paragraph lengths \parencite[]{chakrabortyAIGeneratedText2023}. These tools are trained on vast datasets of human and AI-generated texts, enabling them to identify subtle 'tells' of AI involvement. Unlike traditional plagiarism detectors that compare submissions against a database, AI detectors focus on linguistic and stylistic cues to differentiate human from machine-generated text.

However, the effectiveness of AI detection mechanisms tends to be hampered by continuous advancements in the area. According to \textcite{perkins_genai_detections}, newer models, such as Anthropic’s Claude 3 Opus, generate texts that look very much like human writing, thereby reducing the predictability on which detection tools rely. They also pointed out that the overall precision of algorithms for the recognition of AI-generated content stands at 39.5\% only but decreases to 22\% in the case of adversarial methods. This means that there is a high frequency of false positives where texts written by people are mistaken as having been created by machines. These mistakes pose significant threats, such as unfair treatment of students and the possible unnoticed misuse of genuine AI.

Some of the adversarial techniques that can be employed to circumvent detection by AI tools involve introducing misspellings, writing like non-native speakers, and increasing burstiness in writing styles. These approaches take advantage of weaknesses present in algorithms used to detect AI-generated texts, making them hard to find. For instance, changing sentence structure or introducing typos may imitate human writing patterns, hence confusing discovery software \parencite[]{perkins_genai_detections}.

Additionally, another reason why AI detection tools have limited use is because they are inequitable. Students from wealthier families might use higher-quality (more expensive) AI tools that are able to avoid discovery, a consideration that gives rise to issues of fairness during assessments \parencite[]{sullivanetal_2023}. Moreover, adoption of these systems increases teachers’ workload since ambiguous findings usually need careful reading, and such findings may lead to conflicting interactions with the learners, thereby resulting in exhaustion among educationalists \parencite[]{swiecki_assessment_2022}.

In conclusion, AI detection tools come with certain hitches. These include false negatives and false positives, which may lead to the exclusion of worthy candidates. It is important to consider these issues if we are to make the most of AI detection tools. By doing so, stakeholders would have a higher chance of finding solutions that are not only more effective but also fairer. More importantly, in consideration of these challenges, making GenAI tools prominent in educational assessment processes may promote equal opportunity, minimize educationalists’ workloads in evaluative activities, and enhance testing transparency and validity relative to educational goals \parencite[]{ogunleye_higher_2024}. Therefore, the creation of an AIAS does not only signify a reaction towards technical difficulties brought about by GenAI tools but also implies a way forward towards responsible utilization of its potentials.

\section{Introduction of the AI Assessment Scale (AIAS)}

The AIAS introduced by \textcite{perkins_aias_2024} stands out as the first attempt to systematize the integration of AI in educational assessment. This scale was developed in response to the increasing demand for the introduction of AI tools into education with a view to ensuring academic honesty, instilling ethical practices, and boosting learning outcomes.
AIAS aims at providing a universal benchmark that may be used to measure the extent of AI’s adoption across different levels of educational institutions. It is composed of five tiers, each signifying a specific stage at which AI should be integrated into educational undertakings. The stages are structured in such a manner as to create a pathway for the educationalists to assess their current positions on the use of AI while teaching and learning in order to be informed of future actions.
\begin{itemize}
	\item \textit {Level 1: No AI (Human-Only):} This level represents traditional assessment methods without any AI involvement, ensuring that students rely solely on their knowledge, understanding, and skills.
	\item \textit {Level 2: AI-Assisted Idea Generation and Structuring:} At this level, AI is used to assist in brainstorming and organizing ideas but is not involved in the final content creation.
	\item \textit {Level 3: AI-Assisted Editing:} AI tools are employed to improve the clarity and quality of student-created work, but no new content is generated by AI.
 	\item \textit {Level 4: AI Task Completion with Human Evaluation:} AI completes specific elements of a task with human evaluation, ensuring academic integrity and understanding.
  	\item \textit {Level 5: Full AI Integration:} AI is used extensively throughout the assessment process, collaborating with students to enhance creativity and learning outcomes.
\end{itemize}

The primary purpose of the AIAS is to provide a structured approach to integrating AI in education. By defining clear levels of AI involvement, the scale helps educationalists implement AI tools responsibly and ethically. The benefits of the AIAS include:

\begin{itemize}
	\item \textit {Encouraging Ethical Use of AI:} Being transparent and fair when using AI in education and making sure it lasts long.
	\item \textit {Improving Learning Outcomes:} Using AI for customized learning experiences and immediate responses.
	\item \textit {Keeping Academic Integrity:} Ensuring that AI tools complement educational assessments rather than compromise them.
\end{itemize}

You can see the AIAS from the figure 1 below.

\begin{figure}
	\centering
	\includegraphics[width=0.85\textwidth]{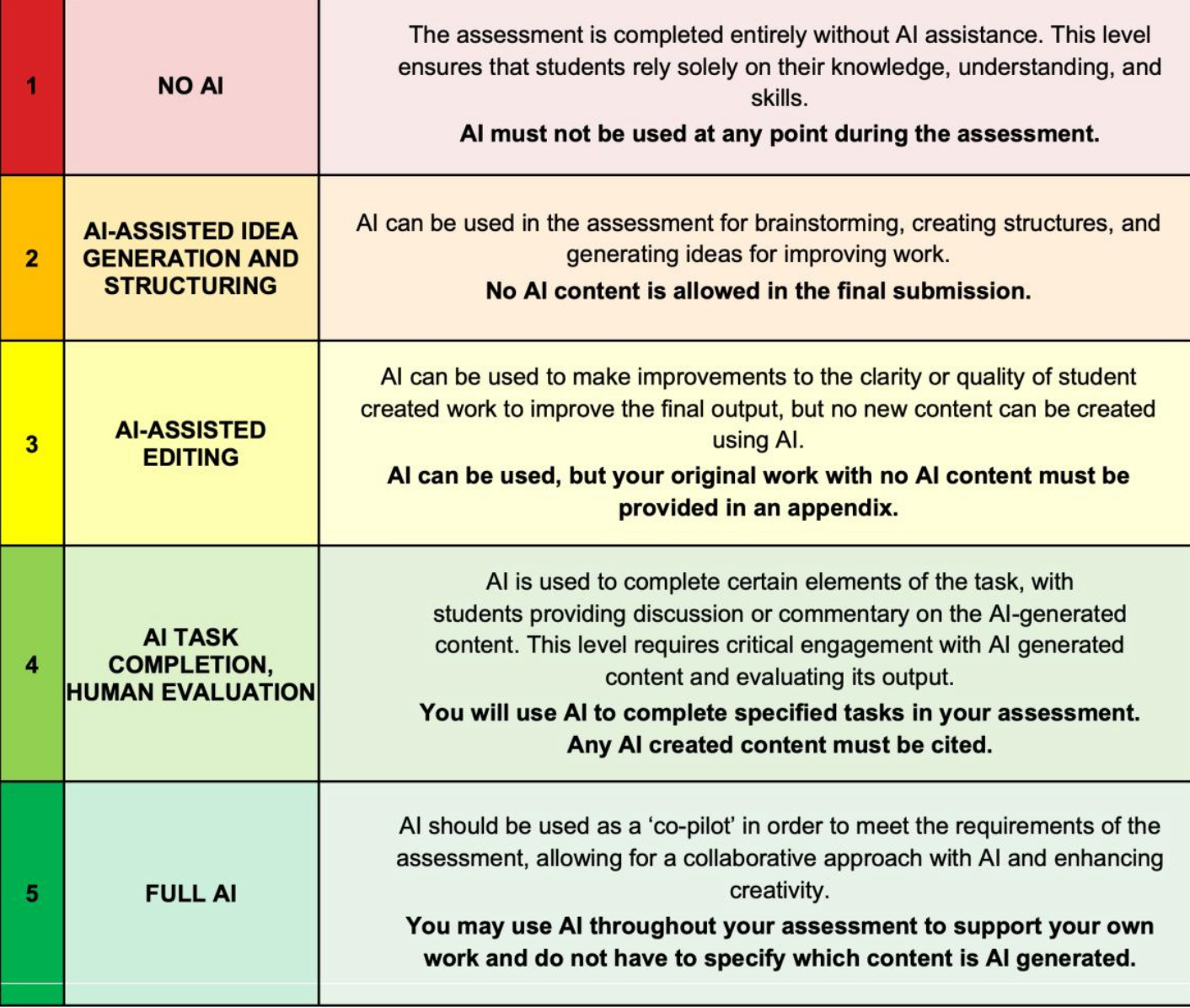}
	\caption{AI Assessment Scale by \textcite{perkins_aias_2024}.}
	\label{fig:aias}
\end{figure}

\section{Transformation of AIAS to Comprehensive AI Assessment Framework (CAIAF)}

Enhancements to the AIAS are essential to address the evolving challenges and opportunities presented by GenAI tools in education. These enhancements include the integration of ethical guidelines, allowance for advanced AI levels, differentiation between educational levels and adjustments, and visual representation and grading variability in AI integration. Each component is crucial for creating a robust, adaptive, and ethical framework for AI use in educational assessments.

\subsection{Ethical Guidelines Integration}
The AIAS is critical to ensuring responsible and beneficial use of AI tools in educational settings. Ethical considerations are paramount in shaping how AI technologies are implemented, addressing issues such as transparency, equity, pedagogical alignment, accountability, and privacy. Using AI without ethical principles would be irresponsible and would lead to significant harm, bias, and inequality \textcite{holmes_ethics_2022}; \textcite{klimova_ethical_2023}.

Aligning these ethical principles with established standards and guidelines from prominent organizations such as the International Society for Technology in Education (ISTE), the Association for Educational Communications and Technology (AECT), the Institute of Electrical and Electronics Engineers (IEEE), and the European Union (EU) ensures a comprehensive framework for the ethical use of technology in education. These standards promote transparency, equity, accountability, and privacy in AI implementation \parencites{jones_doing_2016}{hamiti_ethical_2014}{network_ethical_2001}{mata_ethical_2022}. By embedding these ethical principles into the AIAS, educators can foster an environment that not only makes prominent the benefits of AI technologies but also upholds the highest ethical standards, ensuring responsible and effective AI integration in education. This approach addresses ethical challenges and aligns with greater educational equity, transparency, and accountability goals, as highlighted in recent studies and guidelines \parencites{leimanis_self-imposed_2020}{shih_learning_2021}.

The ethical principles that should be emphasized to each student along with the assignment in a way that is appropriate for the relevant level and guidance on how to achieve them are as follows:

\begin{itemize}
	\item \textit {Transparency:} Students need to declare whether they have used any AI tools or not so that the level of the AI they worked with can be known.
	\item \textit {Equity and Inclusivity:} Equal access to AI technologies should be provided to students of diverse socio-economic backgrounds.
	\item \textit {Pedagogical Alignment:} Employing AI in learning should be geared towards accomplishing educational objectives that promote creativity.
 	\item \textit {Accountability:} The student should ensure the originality of his or her work after using AI, hence not breaching academic ethics.
	\item \textit {Privacy and Data Protection:} AI tools must safeguard student information privacy by upholding data protection laws.
\end{itemize}

\subsection{Integration of Advanced AI Levels with Future Provisions}
Progressive growth in AI technology, especially brought by models like GPT-4o, has made it necessary to add a new stage under the AIAS scale called advanced AI integration (Level 6), which allows for real-time interaction and acts as a personal assistant, indicating the current trendsetter nature of this field.

Level 5 represents a major step for the integration of AI with education in that it relies heavily on the use of various AI tools to aid in different activities during the learning process. At this level, AI tools may be used for content generation, feedback provision, and even assignment grading. However, such applications do not usually provide advanced real-time interaction or highly personalized assistance observed at level 6.

Real-time interaction abilities are what distinguish level 6 from full AI integration through AI tools that are able to answer the students’ questions instantly and organize live tutoring sessions, among other capabilities. AI at level 6 can respond to students’ needs as they study, unlike at level 5, where it may only act within given limits by waiting for the completion of the activities and then responding accordingly later on.

For instance, level 5 AI can create a comprehensive study guide after going through a student’s coursework, while level 6 AI discusses the same guide with the student live by giving immediate answers to any arising questions and by changing the explanations dynamically based on how well or poorly the student is grasping the concepts. This kind of engagement considerably imitates the acts of a human tutor, thus making it possible for learners to get support at the moment they need it.  

While level 5 may have general support tools from AI, level 6 ensures that these supports are customized enough to take into account each student’s unique learning path so far. In the context of contemporary education, it is necessary to create educational experiences that are interactive and flexible since they meet different student requirements and learning styles. More importantly, advanced AI tools have the ability to evaluate a learner’s progress continuously and modify the content accordingly so as to provide them with an individualized learning path.

\subsubsection{Future-Proofing the Framework with Placeholders}
The advancement of AI is no secret, and this is why we need to future-proof our assessment scales by adding placeholders for possible new future levels. This was forecasted just short after the introduction of level 6 due to the rapid development pace, and it is indicative that there may always be something beyond what anyone knows at any given time relevant to such factors. The inclusion of placeholders in advance within such tools as educational instruments meant for AI’s use in educational institutions indicates our understanding of the fast pace of the technology in order to prevent any lagging.

Using placeholders in scales makes them more valuable and reliable in time as they can be easily replaced to reflect the new changes \parencite[]{green_preparing_2020}. This forward-thinking indicates the AIAS’s usefulness for teachers in integrating AI into their teaching methods, even in cases of future technological advancements. A forecast of growth implies that a measure would be constantly updated as technology advances, which therefore makes it relevant within educational setups. For instance, future developments in AI may involve advanced natural language processing abilities, emotional intelligence, or even better adaptive learning algorithms, among others. With the placeholders present in AIAS, it can quickly adapt to and adopt these new advancements so that educationalists have the most recent tools for using AI effectively.

\subsection{Differentiation Between Educational Levels and Adjustments}
The AIAS differentiates between educational levels in order to account for the various stages of development, cognitive abilities, and academic requirements of students at each level. According to \textcite{perkins_aias_2024}, who are the authors of the original study, failure to separate K-12 from higher education was a notable limitation in their work. The primary and secondary education levels were considered to be K-12, while the undergraduate and postgraduate education levels were considered to be higher education. By doing so, specificity was brought into this scale, making it a possible tool for the primary, secondary, and tertiary levels of education. This would facilitate more purposeful, specific adjustments, which could be matched better.

This differentiation makes the scale as specific and detailed as possible, indicating that it is applicable from the beginning to the end of education. It was observed that ethical behaviors developed during childhood tend to persist in adulthood \parencites{badeni_2019}{pushpa_ethical_2012}{puyo_value_2020}{rafikov_prospects_2021}. Thus, this implies that if we adopt such a scale in early childhood programs and then gradually introduce it at higher grade levels, eventually no student would consider it the imposition of anything as they would be accustomed to their own pace \parencite[]{foray_2012}. Likewise, at points where innovation meets resistance among groups who had initially opposed it, there tends become less over time with sustained efforts towards its full integration into systems being seen more as habit forming rather than constituting merely an affront against established routines \parencite[]{ng_innovation_2009}. Consequently, specific adjustments to be developed along with illustrative examples according to different levels of education may be beneficial for the target groups involved.

\subsubsection{K-12 Education}
\paragraph{4.3.1.1.	Primary Education}
The first part of education requires basic knowledge of AI's ethics and safety. AI’s mechanics are learned through the supremacy of practice over theory. According to this method, a child can comprehend all the concepts about AI by performing real acts and easy tasks. In addition, it is important to begin teaching the ethical use of AI as early as possible so that children grow up with responsible attitudes towards technology. Studies show that exposure to ethics at an early stage greatly affects people's long-term behavior and attitudes towards technologies \parencite[]{wang_ethical_2019}.

\paragraph{4.3.1.2.	Secondary Education}
At this level, the learners should be exposed to more complex AI tools as well as their application across different subjects. Similarly, secondary education students need to know its operation principles both at the theoretical and practical level through demonstration of the principles in real-life situations or fieldwork. In addition, attention should be paid continuously to promoting responsible privacy considerations regarding powerful educational use cases for advanced AI technologies. The integration of ethical dimensions into discussions around technology literacy has been identified by some literature works in this domain, and it was found to be beneficial for enhancing students’ comprehension skills associated with critical thinking \parencite[]{pasricha_ethics_2023}.

\subsubsection{Higher Education}
\paragraph{4.3.2.1.	Undergraduate Level}
At this stage, studies focus more on the implementation of things learned at the secondary educational level under STEM domains, especially the ones relevant to practical aspects of AI. Additionally, the social science domains also cannot ignore AI, as it provides some tools through which data analysis may be made much easier and more comprehensible. Therefore, strong emphasis should be placed on integrity as well as responsibility regarding the ethical use of AI. For instance, research has shown that if undergraduate curricula incorporate teachings about AI, significantly better designed programs fostering problem-solving abilities among learners and also enhancing innovational skills may emerge \parencite[]{mollick_using_2023}.

\paragraph{4.3.2.2.	Graduate Level}
AI is used by graduate students in different ways. To apply advanced AI techniques to their specific fields of study, researchers develop complex projects using AI. In order to conduct responsible research, it is important to consider ethics in terms of data integrity and security. According to \textcite{borenstein_emerging_2021}, future challenges will require programs at the graduate level to incorporate advanced applications along with training on ethics.

\subsubsection{Tailored AI Integration Across Educational Levels with Exemplification}
To enhance the scale, distinct examples should be provided for all six levels across various educational settings, such as primary, secondary, undergraduate, and postgraduate. Thus, the tool’s usage will be easier for the educationalists through being more specific about the point of integration of the technology into their curriculum and through the alignment of it with the diverse needs of the students.
Furthermore, if each level has clear examples assigned to them, then the educationalists can gradually understand how AI supports learning at different phases until it is fully implemented without violating any ethical standards, failing which would imply a lack of evidence-based planning guides for assessment purposes, according to \textcite{chan_ai_2023}.

Unique instances at each stage also aid in understanding and following the ethical principles integrated in the AIAS. Through the observation of practical examples for each level, educationalists and learners can gain a better understanding of the ethical concerns and duties related to AI applications at various points. This will make them more compliant with ethical norms in addition to promoting responsibleness and awareness in the use of AI for educational purposes \parencite[]{ma_ethical_2023}.

\subsection{Visual Representation and Grading Variability in AI Integration}

For the AIAS to be effective, there must be ways of visually representing how AI is integrated into various levels. In education, for example, it may be difficult for students and educationalists to understand the extent of the usage of AI without the use of visual tools like color gradients \parencite[]{zhou_designing_2020}. To show the different levels of AI integration graphically, the Comprehensive AI Assessment Framework (CAIAF) uses a range of blue shades from dark to light. Using a red-green gradient would have been wrong because it might mean negative positivity progression which is not all-inclusive. While the red gradient has a negative implication that points to failure or ban, the green gradient, on the contrary, has a positive implication such as success, and this leads to fear or even bias among users, according to \textcite{xu_influences_2023} and \textcite{elliot_color_2014}. A neutral, universally attractive design that fosters clarity and inclusivity visually is created by utilizing this blue spectrum, thus making no reference whatsoever.

With the introduction of “Level 6: Advanced AI Integration,” it became necessary to come up with a new scheme for colored bands. This is because when more levels are added and future improvements embraced, there can only be one continuous-color scale so as not to cause confusion due to many different colors being used concurrently, which could be quite messy in visual terms \parencites{frankel_visual_2012}{singh_creating_2016}{zeileis_escaping_2009}. In addition, such an approach simplifies understanding of the system, thereby preventing cognitive overload among people who may find getting acquainted with complex systems hard enough even without them being represented in visually complex terms, such as multiple distinct hues simultaneously employed.

Educational research supports using visual aids in education in order to improve comprehension and engagement \parencite[]{stobart_developing_2004}. According to studies cited by \textcite{yen_evaluating_2012}, and \textcite{poza-lujan_assessing_2016}; continuous scales along gradients or other similar devices not only help students comprehend difficult concepts better but also increase their interest levels significantly while reducing mental effort during assessments. Therefore, we should adopt an approach that enhances both cognitive ease and appeal within our educational systems through more student-friendly design strategies like these.

\subsubsection{Grading Variability Within Levels}
To accommodate the different degrees of integration at each level, grading variability should be included for enhancing enhance adaptability and effectiveness. This can greatly enhance the adaptability and effectiveness of AIAS. For instance, when it comes to level two (AI-assisted idea generation and structuring), minimal assistance from AI may be provided in brainstorming, whereas extensive support for organizing thoughts may be offered by highly developed systems, depending on the student’s educational stage or grade level. Primary school pupils might use simple tools powered by weak AI that generate basic ideas while they are at the primary educational level, but different software equipped with higher capabilities may come into play during their secondary education so that they can handle topics more holistically.

When it comes to level 3 (AI-Assisted Editing), variability can range from basic grammar and spell checking to complete content revision. In undergraduate education, students could employ AI for the purpose of correcting grammar mistakes and enhancing clarity in writing their essays or reports. At a graduate level, however, such tools may be used to perform extensive editing on academic papers so as to ensure that they meet not only coherence but also the more sophisticated standards expected of such documents.

Moreover, full integration of AI into tasks is allowed at level 5, where learners are free to extensively use these tools throughout their assignments, but this should be reflected in the grading system because different levels of complexity and depth in AI use can warrant various grades. For instance, a student pursuing a bachelor’s degree could employ it for collecting data as well as carrying out preliminary analysis, while another one doing a master’s may integrate them into each stage of research, from hypothesis generation through data analysis to report writing. Hence, even though we are restricted to six points on our scale, there will still be a wide range of potential applications and assessments within each point.

Moreover, the use of a gradient color scheme improves the visual indication of assessment diversity. This is because, in the former scale, each level was represented with a single color, which made someone think that the levels themselves could not be graded, but this approach helps to eliminate such misunderstandings by using different shades of AI integration within each level.

As a result, if the current revisions proposed for the AIAS developed and put into use by \textcite{perkins_aias_2024} are carried out, this version, which is now more inclusive, will have a more accurate orientation and will be more educationalist- and student-friendly by evolving into the one observed. At the same time, this scale, which was formed after the recent revisions, is now named the Comprehensive AI Assessment Framework (CAIAF). This framework can be seen in the Appendix part, and it can also be \href{https://www.selcukkilinc.com/_files/ugd/929438_d0780064229c470f9cd1fde7f519b461.pdf}{downloaded for a full-screen view}. 

\section{Conslusion and Suggestions}

This article examines how GenAI tools have changed education by showing the opportunities and challenges they bring. The reactions against the use of AI in educational assessments are diverse, and dealing with these reactions has been at the center of attention. Problems that come with integrating AI were investigated and answered through the identification of the need for ethical principles and the determination of different educational levels, advanced levels of AI, and visual representations, among others.

A significant part of this study focused on improving the original AIAS created by \textcite{perkins_aias_2024}. This was achieved by adding ethical guidelines, introducing higher levels of AI, and ensuring clear differentiation between K-12 and higher learning institutions. Additionally, a multi-colored grading system with varying degrees has been adopted for more detailed results, thereby leading to the development of the CAIAF, which provides strong yet flexible guidelines for the ethical use of AI in educational settings. 

This work has deep future implications for educational practices. With the continuous advancement of AI, its incorporation into the system should be done cautiously so as to maximize its benefits while mitigating its risks. CAIAF lays down ground rules but still needs more improvements, and adjustments can be made depending on various contexts. Future studies may try out this model in different settings with the aim of gauging its efficacy and pointing out areas that may require modifications.

Furthermore, in order to adapt to the ever-changing landscape of AI, the model has to be flexible enough to incorporate new advancements and applications. This means that as the world becomes more AI-oriented, our teaching methods should not only be up-to-date but also efficient. Placeholder values at different levels of integration for future use within formal education underscore the necessity for continuous improvement.

For effective implementation and utilization of the holistic AI assessment model, it is recommended that educationalists, policy-makers, and members of academia embrace, use, and improve the framework regularly. There is a need for cooperation among these stakeholders to ensure that AI is applied in an ethical and constructive manner in schools.

It is important to note the desire for continuing discourse and exploration. One must be prepared at all times for such changes brought about by the rapid advancements in AI in order not to fall behind where educational practice is concerned. By cultivating a climate where change is always welcomed and considered as a chance to grow –instead of shunning new ideas-, we can use AI to improve our teaching methods significantly.

However, the primary focus for future work should be around implementing this framework in real educational settings. Through its practical application, we hope to gain much-needed insight into what works or fails and, thus, to further shape our model accordingly. It is therefore incumbent upon us, as educationalists, to be committed to advancing learning experiences through AI integration within schools, colleges, universities, etc., and to continue using this instrument rigorously so as to ensure that there is continuous improvement in relation to the effectiveness of different AI systems meant for supporting various aspects involved in the education sector.

\printbibliography

\begin{appendices}
\section{Comprehensive AI Assessment Framework}

    \raggedright
    \includegraphics[width=800px,height=620px,keepaspectratio]{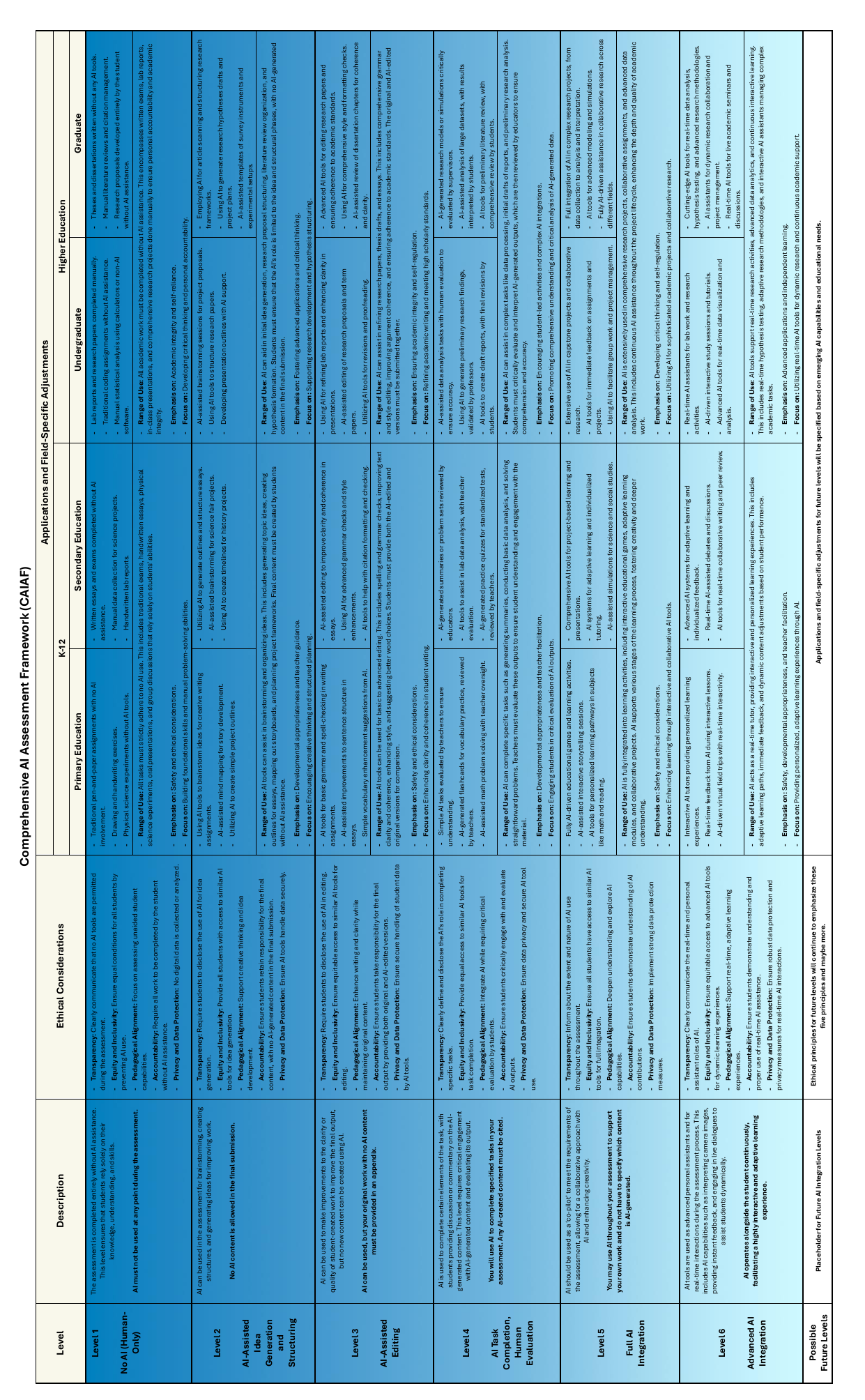}

\end{appendices}
\end{document}